\documentclass[aps,prl,twocolumn,groupedaddress,showpacs,nofootinbib,
nobibnotes,twoside]{revtex4}
\usepackage{graphicx} 
\usepackage{hyperref}
\usepackage{latexsym}
\usepackage{textcomp}
\usepackage{epsfig}

\newcommand\fverb{\setbox\pippobox=\hbox\bgroup\verb}
\newcommand\fverbdo{\egroup\medskip\noindent%
            \fbox{\unhbox\pippobox}\ }
\newcommand\fverbit{\egroup\item[\fbox{\unhbox\pippobox}]}
\newbox\pippobox

\def\tdm{\tilde{R}}
\def\tdr{\tilde{R}}
\newcommand{\gsim}{\begin{array}{c}\sim\vspace{-21pt}\\> \end{array}}

\preprint{FERMILAB--PUB--05/381--T\\EFI--05--12}
\catcode`\@=11
\def\ps@fnal{\def\@oddhead{\textsf{FERMILAB--Pub--05/381--T \hfil \thepage}}
\def\@evenhead{\thepage \hfil \textsf{FERMILAB--Pub--05/381--T}}}
\relax

\begin{document}

\pagestyle{fnal} %%% FOR PREPRINT
%
%\phantom{M} \hfill \textsf{FERMILAB--PUB--05/381--T}\\[-48pt]

%\preprint{}

\title{Randall-Sundrum with AdS(7)}


\author{Ruoyu Bao}
%\thanks{}
\affiliation{Enrico Fermi Institute and Department of
Physics,\\ University of Chicago, Chicago, IL 60637, USA}

\author{Joseph Lykken}
\affiliation{Theoretical Physics Department, Fermi National 
Accelerator Laboratory,\\ P.O.\ Box 500, Batavia, IL 60510 USA}


%\date{\today}

\begin{abstract}
In the same sense that $AdS_5$ warped geometries arise naturally 
from Type IIB string theory with stacks of $D3$ branes, $AdS_7$ warped 
geometries arise naturally from $M$ theory with stacks of $M5$ branes. 
We compactify two spatial dimensions of $AdS_7$ to get $AdS_5\times\Sigma^2$,
where the metric for $\Sigma^2$ inherits the same warp
factor as appears in the $AdS_5$.
We analyze the 5d spectrum in detail for the case
of a bulk scalar or a graviton in  $AdS_5\times T^2$, in a setup
which mimics the first Randall-Sundrum model. The results display
novel features which might be observed in experiments
at the CERN Large Hadron Collider. For example, we obtain TeV scale string winding states
without lowering the string scale. This is due to the {\it double warping} 
which is a generic feature of winding states along compactified $AdS$
directions. Experimental verification of these signatures of $AdS_7$ 
could be interpreted as direct evidence for $M$ theory.
\end{abstract}

\pacs{11.25.Mj, 11.25.Uv, 11.25.Yb, 11.10.Kk \hfill 
\fbox{\textsf{FERMILAB--PUB--05--381--T}}} 
%%%%%%%%%%%%%%%%%%%%%%%%%%%%%%%%%%%%%%%%%%%%%%%%%%%%%%%%%%%%%%%%%%%%%%%%%%%%%%%%%
%                                                                               %
%   98.80.Cq Particle-theory                                                    %
%   and field-theory models of the early Universe (including cosmic             %
%   pancakes, cosmic strings, chaotic phenomena, inflationary universe,         %
%   etc.)                                                                       %
%   98.70.Vc Background radiations                                              %
%   98.65.Dx Superclusters; large-scale structure of the Universe               %
%   95.35.+d Dark matter (stellar, interstellar, galactic, and cosmological)    %
%\keywords{}

\maketitle


%%% ----------------------------------------------------------------------

\section{Introduction}

The five-dimensional warped models of 
Randall and Sundrum \cite{Randall:1999ee} can be
thought of as stripped down effective field theory approximations
to the near horizon geometry \cite{Aharony:1999ti}
of a stack of $N$ parallel $D3$ branes
in Type IIB string theory. This near horizon geometry is
$AdS_5\times S^5$, where the Anti-de Sitter radius $R_{AdS}$ and
the radius of the sphere are equal, scaling like $N^{1/4}$.
Thus for large $N$ the $AdS$ physics involves an energy scale
that is parametrically lower than the string (or Planck) scale.
So does the physics on $S^5$, but we also expect this physics to
be drastically altered as one breaks supersymmetry, and otherwise
attempt to make the compactification more realistic. Since in
any event the sphere $S^5$ is not warped, in seems reasonable to
factor out this physics and examine low energy effective 5d field theory
on $AdS_5$. In the Randall-Sundrum (RS) approach, a phenomenological
3-brane called the Planck brane is added to truncate $AdS_5$ in
a way that excludes the $AdS$ boundary. From the point of view of
string theory, this is a boundary condition that replaces the
original matching of the near horizon geometry to $M_5\times S^5$;
in the $AdS/CFT$ dual language, the Planck brane is a particular
choice of UV cutoff for the 4d $CFT$ coupled to 4d gravity. The
first Randall-Sundrum model (RS1) also introduces a second brane,
called the TeV brane. This brane truncates the $AdS$ geometry in
the other direction, excluding now the $AdS$ horizon. This is a
wise idea for a low energy effective field theory approach, since
close to the horizon the field theory cutoff necessarily shrinks
to zero and stringy effects become important.

In this paper we examine an analogous story which starts with
eleven dimensional $M$ theory. The low energy supergravity theory has a
solution \cite{Aharony:1999ti}
corresponding to a stack of $N$ parallel $M5$ branes.
These $M5$ branes are just the magnetic sources for the
antisymmetric tensor gauge field $A_{MNP}$ of 11d supergravity,
which also has electric sources called $M2$ branes. We
extract the near horizon geometry of this solution
for (moderately) large $N$. This geometry is
$AdS_7\times S^4$, where
$R_{AdS} = 2(\pi N)^{1/3}\ell_p$,
with $\ell_p$ the Planck length, in the convention where the
$M2$ branes have inverse tension $(2\pi )^2\ell_p^3$.

As motivated above we will discard the unwarped sphere $S^4$
and consider low energy effective 7d field theories built on
$AdS_7$:
\begin{eqnarray}
ds^2 = a^2(y)\left( 
\eta_{\mu\nu}dx^{\mu}dx^{\nu}
+\delta_{ij}dx^idx^j\right)
+dy^2 \; ,
\label{eqn:fdef}
\end{eqnarray}
where we have introduced the warp factor $a(y) = {\rm exp}(-ky)$,
$k = 1/R_{AdS}$.
Our metric signature convention is $(-++++++)$.
As in RS, we have to tune the brane tensions to obtain flat
6d slices.
Note that $AdS_7$ setups are very restricted by the requirement
of anomaly cancellation in the 
6d boundary gauge theories \cite{Appelquist:2002ft}.

\section{$\mathbf{AdS_5}$ with extra warpings}

To make contact with a Randall-Sundrum type construction, we
now compactify two spatial dimensions of the $AdS_7$.
Thus 
\begin{eqnarray*}
AdS_7\rightarrow AdS_5\times\Sigma^2 \quad .
\end{eqnarray*}
The simplest choice for $\Sigma^2$
is a torus $T^2$, since
in this case we do not need any additional sources in the
Einstein equations to get a consistent background solution.
The metric becomes:
\begin{eqnarray}
ds^2 = a^2(y)
\eta_{\mu\nu}dx^{\mu}dx^{\nu}
+a^2(y)R^2\left(d\theta_1^2+d\theta_2^2\right)
+dy^2 \; ,
\end{eqnarray}
where for simplicity we assume a common radius parameter $R$.
Note that the measured radius of the torus includes
a warp factor.

Almost as simple is a compactification on a sphere $S^2$.
Since the sphere has curvature, this requires an
additional bulk source. As discussed for example in
\cite{Gherghetta:2000jf},
two bulk scalar fields with a simple
potential allows a solution of the equations of motion
with a ``hedgehog'' profile for the scalars.

Any 7d bulk field can be expanded in a complete orthonormal
basis of Kaluza-Klein (KK) modes for $\Sigma^2$. These are just
sines and cosines for $T^2$, and spherical harmonics for $S^2$.
Substituting the KK expansion into the 7d action, we can
then integrate over $\Sigma^2$ to obtain an effective 5d action.

We will do this explicitly for the case of a 7d bulk scalar
on $AdS_5\times T^2$, and then extend the results to discuss
the 7d graviton. For the torus with radius $R$ the squared KK momenta
are
\begin{eqnarray}
g^{ij}p_ip_j = {n_1^2 +n_2^2\over a^2R^2} \equiv {\vec{n}^2\over a^2R^2}
\; ,
\end{eqnarray}
where $n_1$, $n_2$ are integers.

In the underlying theory there will in general be additional bulk
modes arising from $M2$ or $M5$ branes wrapping one or more cycles
of the torus. These will be particle-like modes if the the other
dimensions of the branes are already wrapped around compact cycles
of $S^4$, or whatever replaces $S^4$ in a realistic model. For example,
an $M5$ brane wrapped on $S^4$ 
is a baryonic string \cite{Witten:1998xy}, which can
in turn wind around cycles of the torus. Such modes can be
labelled by their winding numbers, which form a two-vector
of integers $m^i$, $i =1$,2.

Whereas the KK mode contributions to the
world-sheet Hamiltonian of the string are proportional to
$\ell_p^2g^{ij}p_ip_j$, the winding modes contribute
\begin{eqnarray}
{R^2\over\ell_p^2}g_{ij}m^im^j = {R^2\over\ell_p^2}a^2\vec{m}^2 \quad .
\end{eqnarray}
Note that the parameter $\ell_p$ above will in general
differ from the $\ell_p$ introduced above (\ref{eqn:fdef}),
by a factor of order one which depends upon how strings
emerge from the original 11d $M$ theory.

We can write down the equations of motion for the bulk
scalar in the 5d effective theory, expanded into KK and
winding modes $\phi_{\vec{n}\vec{m}}(x^{\mu},y)$:
\begin{eqnarray}
\left[{1\over a^4}{d\over dy} a^6{d\over dy}
- p^2 + {\vec{n}^2\over R^2} 
-a^2m_b^2 
- a^4{\vec{m}^2R^2\over \ell_p^4}\right]\phi_{\vec{n}\vec{m}} = 0 \, ,
\label{eqn:seom}
\end{eqnarray}
where $m_b$ is an explicit bulk mass from the 7d theory.

The usual warping of Randall-Sundrum is visible in (\ref{eqn:seom})
as the fact that the bulk mass-squared term is multiplied by
$a^2(y)$ relative to the 4d momentum-squared $p^2=p_{\mu}p^{\mu}$.
We observe as well two new kinds of behavior:
\begin{itemize}
\item The KK mode mass-squared contributions
from the torus have no warping relative
to the 4d momentum-squared $p_{\mu}p^{\mu}$.
\item The winding mode mass-squared contributions have double warping
compared to the ordinary bulk mass term, \textit{i.e.,} they are
multiplied by $a^4(y)$ relative to the 4d momentum-squared $p_{\mu}p^{\mu}$.
\end{itemize}

Now we change variables to the conformal coordinate $z =1/ka(y)$,
suppress indices, and perform the useful rescaling $\phi(z) = k^3z^3h(z)$.
The equation of motion becomes:
\begin{eqnarray}
z^2h^{\prime\prime} + zh^{\prime}
+\left[(\mu^2 - {\vec{n}^2\over R^2})z^2
-\nu^2 -{\vec{m}^2R^2\over k^4\ell_p^4}{1\over z^2}
\right]h = 0 \, ,
\label{eqn:fseom}
\end{eqnarray}
where $\nu^2 = 9+m_b^2/k^2$, and we are
putting the 4d momenta on-shell, with
$\mu^2 = -p_{\mu}p^{\mu}$ denoting the physical 4d mass.
It is the mass spectrum of $\mu$ that we want to determine
from an analysis of the solutions of (\ref{eqn:fseom}).
For $\vec{n}^2 = \vec{m}^2 = 0$, this is the equation of motion
for a 5d massive bulk scalar in $AdS_5$, as first analyzed in
\cite{Goldberger:1999wh}.
In general (\ref{eqn:fseom}) is equivalent to
the Mathieu equation \cite{Mathieu}.

\section{5d Kaluza-Klein mode analysis}

We can take (\ref{eqn:fseom}) as the starting point for a
5d Kaluza-Klein mode analysis in $AdS_5$. To get the
analog of scalars in a Randall-Sundrum model, we can impose Neumann
boundary conditions at the location of one or two flat branes.
These branes have co-dimension one, i.e. they are 5-branes, located
at $z=1/k$ and $z=1/T$. The parameter $T$ is assumed to be of
order the TeV scale.

To analyze graviton modes, we use the fact that in \textit{e.g.} harmonic
gauge each component of the graviton obeys (\ref{eqn:fseom}).
The appropriate boundary conditions for the graviton modes,
after tuning the brane tensions to their conventional RS
values, are:
\begin{eqnarray}
\left[zh'(z)+3h(z)\right]_{z=1/k} = \left[zh'(z)+3h(z)\right]_{z=1/T} = 0\; .
\end{eqnarray}
This means that the graviton mode solutions are identical to 
the $m_b = 0$ scalar solutions, up to an overall factor.

Let us start with the simple case where $\vec{m}=0$.
The solution to (\ref{eqn:fseom}) 
depends upon the sign of
$\mu^2 - \vec{n}^2/R^2$. There are no solutions
for $\mu^2 - \vec{n}^2/R^2 < 0$. For
$\mu^2 - \vec{n}^2/R^2 > 0$
the solution is
\begin{eqnarray}
\textstyle{
\phi(z)={k^3z^3\over N}
\left[J_\nu\Big(z\sqrt{\mu^2-{\vec{n}^2\over R^2}}\Big)
+bY_\nu\Big(z\sqrt{\mu^2-{\vec{n}^2\over R^2}}\Big)\right].
}
\label{eqn:solsa}
\end{eqnarray}
If we only consider the case of light modes where 
$\mu^2-{\vec{n}^2\over R^2}\ll k^2$,
then the boundary conditions determine $b$ and yield
\begin{eqnarray}
3J_\nu(x)+xJ_\nu^\prime(x)=0\; ,
\label{eqn:specsa}
\end{eqnarray}
where $xT=\sqrt{\mu^2-{\vec{n}^2\over R^2}}$.
The roots of (\ref{eqn:specsa}) determine the mass spectrum.
These solutions represent towers of
toroidal KK modes on top of the TeV-spaced
massive warped KK modes of Randall-Sundrum. The RS spectrum itself
is also modified, since \textit{e.g.,} for $m_b=0$, $\vec{n}=0$ the relevant
Bessel functions are $J_3(\mu z)$ and $Y_3(\mu z)$, rather than
$J_2(\mu z)$ and $Y_2(\mu z)$ as in standard RS.

Since the toroidal KK modes are not warped, $R\gsim 1/T$ corresponds
to compactifications large enough that we will see the toroidal KK excitations
with spacings comparable to the Randall-Sundrum KK modes. In order not
to reintroduce the hierarchy problem, we would then need some mechanism
that naturally stabilizes $R$ at inverse TeV values. By contrast,
if our torus were compactifying flat rather than $AdS$ directions, 
then $R\sim 1/k$ would give TeV-spaced KK excitations \cite{Davoudiasl:2002wz}.
 
In the special case
$\mu^2 - \vec{n}^2/R^2 = 0$,
the solution to the equation of motion is
\begin{eqnarray}
\phi(z)={k^3z^3\over N}\left( z^{-\nu}+bz^{\nu}\right)\; .
\label{eqn:solsb}
\end{eqnarray}
Similarly, the boundary conditions give us
\begin{eqnarray}
(3-\nu)k^\nu + b(3+\nu)k^{-\nu}&&=0\; ,\\
(3-\nu)T^\nu + b(3+\nu)T^{-\nu}&&=0\; .
\label{eqn:bcsb}
\end{eqnarray}
This set of equations has no solution except when $\nu =3$,
\textit{i.e.,} when there is a scalar zero mode.
The solutions then represent the tower of toroidal
KK modes on top of the zero mode.

\section{5d winding mode analysis}

The most interesting new feature of our analysis is the
spectrum of winding modes. 
For simplicity, we will only consider the simplest case,
with $\vec{n}=0$ and $m_b=0$.
Then (\ref{eqn:fseom}) reduces to:
\begin{eqnarray}
z^2h''+zh'+\Big(\mu^2z^2-9-{\tilde{R}^2\over z^2}\Big)h=0\; ,
\label{eqn:maineq}
\end{eqnarray}
where we have introduced the notation
$\tilde R=R\vert\vec{m}\vert/k^2 \ell_p^2$.
As usual in Randall-Sundrum setups, we will assume that
$k\ell_p$ is less than 1 but not dramatically so, \textit{e.g.,}
$k\ell_p \simeq 0.1$.

First of all, we want to see if there exists a zero mode.
Setting $\mu=0$ the solution of (\ref{eqn:maineq}) is
\begin{eqnarray}
\phi(z)={k^3z^3\over N}\left[
I_\nu\Big({\tilde{R} \over z}\Big)
+bK_\nu\Big({\tilde{R}\over z}\Big)
\right]
\; .
\label{eqn:zerosol}
\end{eqnarray}
The boundary conditions give:
\begin{eqnarray}
b&&=-{3I_\nu(k{\tilde{R}})
-{k\tilde{R} }\,
I_\nu^\prime(k{\tilde{R}})
\over
3K_\nu(k{\tilde{R}})
-{k\tilde{R} }\,
K_\nu^\prime(k{\tilde{R}})
} = [ k \leftrightarrow T ] \; ,
\end{eqnarray}
which cannot both be true at the same time unless $T=k$. 
Therefore there is no zero mode for the case with winding modes only.

For the massive winding modes, the solutions can be found numerically.
Here we summarize the results.

\subsection*{$\mathbf{R\sim 1/k}$}

This corresponds to the smallest compactification radius, where
we expect the winding modes to be the lightest. Like the KK modes,
these should appear as additional excitations on top of the TeV-spaced
warped KK modes of Randall-Sundrum. We find that the
RS zero mode does not have any winding excitations. The
winding excitations of the lowest lying massive RS modes have
masses given by $\mu=\mu_1 - 0.40\,\tdr^2T^3$ and
$\mu = \mu_2 +1.46\,\tdr^2T^3$, where $\mu_1=5.13\,T$
and $\mu_2 = 8.42\,T$ are the masses of the first two
massive RS modes. 
Notice that the mass splittings are very small,
of order $10^4\times T^3/k^2 \sim 10^{-12}$ eV.

\subsection*{$\mathbf{R\sim 1/T}$}

\begin{figure}[htbp]
  \begin{center}
    \psfig{file=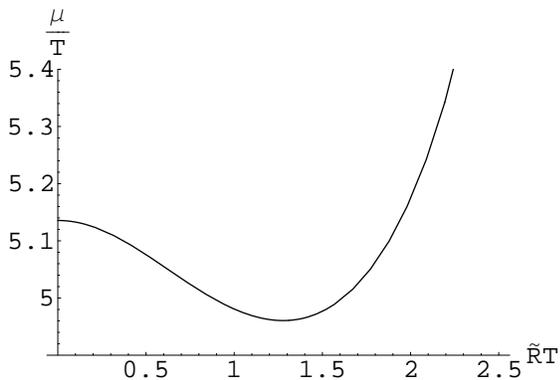}
    \caption{Winding excitations of the lightest RS massive mode, for 
$\tdm$ of order $1/T$.}
    \label{fig:1}
  \end{center}
\end{figure}

This corresponds to a large compactification radius, thus were it
not for warping we would expect the winding excitations to be
extremely heavy. However, because of the double warping of the
winding term in (\ref{eqn:fseom}), we find that in this case the
winding excitations are actually of order T. Once again
the RS zero mode does not have any winding excitations.
The winding excitations of the lowest lying massive RS mode have
masses as shown in Figure \ref{fig:1}. Note that the spacing of the
winding modes can be considerably less than $T$ if $R$ is small,
\textit{e.g.,} for $RT \sim 0.01$ and $k\ell_p \simeq 0.1$ the spacing
is a fraction of $T$.

\section{Double warping versus T-duality}

Our results for the spectrum of KK and winding modes can be understood
qualitatively as a combination of warping with the usual T-duality
relation between the KK and winding spectra. For simplicity we will
not distinguish between $k$ and $\ell_p$ in this discussion.

When the compactification
radius $R$ is of order $1/T$, the toroidal KK modes are or order $T$,
with no additional warping. The T-dual winding modes are of order $k^2/T$,
multiplied by a double warp factor $T^2/k^2$; thus the winding modes
are of order $k^2/T\times T^2/k^2 \sim T$, as reported above.

When the compactification
radius $R$ is of order $1/k$, the toroidal KK modes are of
order $k$, with no additional warping. The T-dual winding modes
are of order $k$ as well, but again multiplied by a 
double warp factor $T^2/k^2$. The winding excitations are on top
of the massive warped KK excitations, which are of order $T$.
Since the masses add in quadrature, we expect the splittings of
the winding mode excitations to be of order
\begin{eqnarray}
\sqrt{T^2 + {T^4\over k^2}} - T \sim {T^3\over k^2} \; ,
\end{eqnarray}
as reported above.

\section{LHC discovery opportunities}

The $AdS_7 \to AdS_5\times\Sigma^2$ scenarios presented here are as
well-motivated as the $AdS_5$ effective theories that are the
basis for standard Randall-Sundrum. Both are stripped down versions
of physics that could actually emerge from string theory in a
robust way, albeit decorated with many stringy complications.

The spectrum of bulk graviton and scalar
modes in our $AdS_7$ models can differ from standard RS1 in several ways.
First of all, because we start with $AdS_7$ instead of $AdS_5$,
the spacing of the massive graviton modes is determined by the
zeroes of the Bessel function $J_2(\mu T)$, rather than
$J_1(\mu T)$. The smoking gun signature of RS1 at the LHC 
is \cite{Davoudiasl:1999jd}
a tower of resonances with masses 1, 1.83, 2.66, 3.48, ..., in units
of the first heavy resonance. For our models, the prediction is
a tower of resonances with masses 1, 1.64, 2.26, 2.88, ..., in units
of the first heavy resonance.

LHC experiments will also be sensitive to the extra toroidal KK states,
if $R$ is of order a TeV. The spectrum of these additional excitations
is evenly spaced, distinguishing it from extra toroidal dimensions
not associated with $AdS$ directions \cite{Davoudiasl:2002wz}.

LHC experiments may be sensitive to our extra winding states, depending
upon which bulk modes have winding excitations, and how they couple
to Standard Model particles. For $R$ of order $1/T$,
the winding states have discrete mass spacings
or order $T$. For $R$ of order $1/k$, the winding states
are essentially a continuum on top of a mass gap of $5.13\,T$. This
is somewhat reminiscent of LR models \cite{Lykken:1999nb}, but clearly
distinct since in that case there is no mass gap in the continuum
KK spectrum.

Light winding states have been discussed in the 
literature \cite{Donini:1999px} in the
context of 
ADD models \cite{Arkani-Hamed:1998rs} with a TeV 
string scale \cite{Lykken:1996fj}.
Here we have obtained TeV scale string winding states
without lowering the string scale.
This is due to the double warping which is a
generic feature of winding states along compactified $AdS$
directions.

\subsection*{Acknowledgments}
The authors are grateful to Bill Bardeen, Bogdan Dobrescu, Minjoon Park,
and Daniel Robbins for useful discussions. JL acknowledges the support
of the Aspen Center for Physics.
This research was supported by the U.S.~Department of Energy
Grants DE-AC02-76CHO3000 and DE-FG02-90ER40560.

% ------------------------------------------------------------------------

\end{document}